\documentclass[%
 reprint,
 amsmath,amssymb,
 aps,
]{revtex4-2}

\usepackage{graphicx}
\usepackage{dcolumn}
\usepackage{bm}
\usepackage{color}
\usepackage{physics}
\usepackage{placeins}
\usepackage{caption}
\usepackage{lineno} 

\usepackage[colorlinks=true, allcolors=blue]{hyperref}

\makeatletter
\renewcommand{\fnum@figure}{\textbf{Figure \thefigure}}
\makeatother

\usepackage[font=small,labelfont=bf,
   justification=Justified,
   format=plain]{caption}

\begin{document}


\preprint{APS/123-QED}

\title{Epistatic pathways in evolvable mechanical networks}

\author{Samar Alqatari}
 \email{samarq@uchicago.edu}
\author{Sidney R. Nagel}%

\affiliation{%
 Department of Physics and The James Franck and Enrico Fermi Institutes \\ 
 The University  of  Chicago, 
 Chicago,  IL  60637,  USA.
}

\date{\today}

\begin{abstract}
An elastic spring network is an example of evolvable matter. It  can be pruned to couple separated pairs of nodes so that when a strain is applied to one of them the other responds 
either in-phase or out-of-phase. 
This produces two pruned networks, with incompatible functions,
that are nearly identical but differ from each other by a set of “mutations” each of which removes or adds a single bond in the network. 
The effect of multiple mutations is epistatic, that is, 
the effect of a mutation depends on what other mutations have already occurred. We generate ensembles of network pairs that differ by a fixed number, $M$, of discrete mutations and evaluate all $M!$ mutational paths between the in- and out-of-phase behaviors up to $M=14$.  With a threshold response for the network to be considered functional, so that non-functional networks are disallowed, only some mutational pathways are viable. 
We find that there is a surprisingly high critical response threshold above which no evolutionarily viable path exists between the two networks. The few remaining pathways at this critical value dictate much of the behavior along the evolutionary trajectory. In most cases, the mutations break up into two distinct classes.  The analysis clarifies how the number of mutations and the position of a mutation along the pathway affect the evolutionary outcome. 
\end{abstract}

\maketitle

The behavior of matter -- both living and non-living -- stems from the physical interactions between its constituents. In evolutionary biology, epistatic interactions are those where the effect of one local change in the genetic code strongly depends on the precise set of preceding mutations; that is, the effects of multiple mutations on an organism do not simply add~\cite{phillips2008epistasis, Breen2012primary, poelwijk2016context, starr2016epistasis, adams2019epistasis}. This is important because evolutionary pathways can become non-viable if the mutations occur in an unfavorable order~\cite{weinreich2005perspective, Weinreich2006darwinian, poelwijk2007empirical,de2014empirical, ferretti2018evolutionary, poelwijk2019learning}. A purely mechanical network provides a platform where evolvable matter can be  thoroughly examined to provide a deeper understanding of the processes governing the transformation of function. Here we show that such networks are strongly epistasic and use them to recapitulate the behavior and elucidate the mechanisms of biological evolution; single mutations can act as switches between two highly fit variants with incompatible functions~\cite{maynard1970natural}. We discover that mutations systematically fall into one of two distinct epistatic classes. This inherent structure makes evolutionary trajectories more predictable and produces an effective memory of the two variants' common ancestor. Epistasis in purely physical systems provides a unique platform where ideas about evolvable matter from physics and biology can coexist and take inspiration from one another. 

We study disordered elastic mechanical networks that have been previously used to replicate aspects of protein behavior~\cite{atilgan2001anisotropy, hemery2015evolution, rocks2017designing, yan2017architecture,tlusty2017physical, flechsig2017design, bravi2020direct}. In those studies, long-range coupling,  mimicking allostery in proteins~\cite{perutz1989mechanisms,gunasekaran2004allostery}, could be tuned into a mechanical network by pruning bonds in such a way that applying a local strain at a “source” triggers a desired strain response at a distant “target”~\cite{rocks2017designing,yan2017architecture, bravi2020direct, rouviere2023emergence}. 
It was demonstrated that programming allostery into mechanical networks is surprisingly straightforward and nearly any two source and target pairs can be coupled. 

\begin{figure}[phtb!]
\centering
\includegraphics[width=\linewidth]{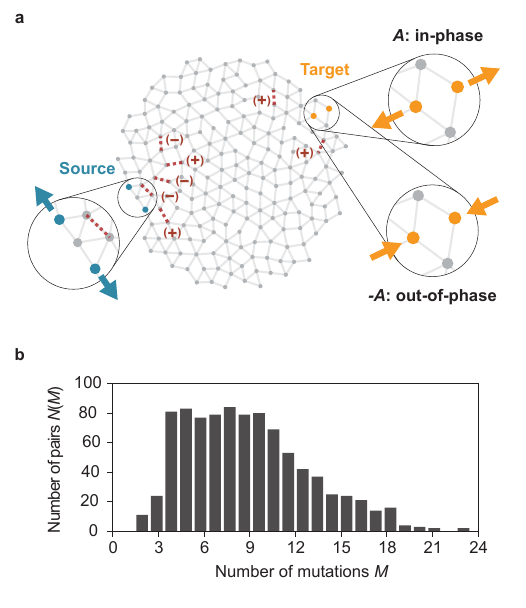}
\caption{\textbf{Ensemble of source-target allosteric pairs tuned for function-switch in a mechanical network.} \textbf{a}. Example sites, designated “source” and “target”, tuned separately for in-phase $A$ and out-of-phase $-A$ strain response, as shown. 
In this example, the $M=7$ mutations, 
consisting of either bond additions $(+)$ or removals $(-)$ highlighted in red, result in a $180^\circ$ change of phase at the target. 
The network shown has 200 nodes and 421 bonds. 
\textbf{b}. Distribution of number of mutations $N(M)$ to switch response for 910 sampled source-target pairs in the same network, each tuned for strain ratio $\eta^* = 1.0$.}
\label{fig1_network}
\end{figure}

In addition, if two nodes at the source are pulled apart, the system can be tuned so that the pair of nodes at the target 
either responds in-phase or else out-of-phase with the source, 
as shown in Fig.~\ref{fig1_network}a. Thus, just as a network can be tuned to perform a specific function, which we call $A$, by pruning one set of bonds, it can be separately tuned to perform the negative of that function, $-A$, by removing another independent set.

Clearly a system cannot be tuned for both $A$ and $-A$ simultaneously and the sets of bonds that are removed in the two cases are not identical. However, the network can be made to switch between the two incompatible functions by performing a set of discrete modifications. For example, 
one can create an evolutionary trajectory from $A$ to $-A$ by restoring the bonds that had been removed to produce $A$, as well as deleting those that produced $-A$. These bond
additions and/or removals are an analog of biological “mutations” that allow the network to evolve between the two functions. The mutations can be applied in any order.

Such a physical network demonstrates the effects of mutations on an evolutionary fitness landscape~\cite{wright1932roles,gavrilets2004fitness,de2014empirical}. It only deals with mechanical interactions yet, importantly, it is \textit{not} over-simplified; it represents an accurate and experimentally validated model of a physical system with all its complexities and interactions between components~\cite{rocks2017designing} and a well-defined measure of function fitness. It provides an excellent platform for studying the evolvability of matter, and structure of epistasis, abstracted from any biochemical complexity.

Because the network tuning algorithms are so effective at coupling pairs of nodes for either in- or out-of-phase behavior~\cite{rocks2019limits}, we can create and evaluate large ensembles of function switches which is difficult to do in a purely biological context~\cite{kleeorin2023undersampling}. 
This allows us to examine evolution for many examples that have the same number of mutations, $M$, between pairs of incompatible functions. We examine individual examples of how mutations lead to function switch and characterize statistically the observed mutational pathways.

We use these physical networks to investigate evolutionary behavior. In particular, in a biological context some pathways can be inaccessible because at some point the system loses all function~\cite{maynard1970natural, poelwijk2007empirical}. In our case, this corresponds to being below the threshold for adequately performing either function
$A$ or function $-A$. We evaluate the pathways that become blocked as a function of the fitness threshold value. Above a critical value, no function-preserving pathways remain. We find this critical value to be surprisingly high.

\section{\label{sims}Mechanical network allostery}

\subsection{Tuning mechanical networks for incompatible allosteric functions}

Starting with a network with coordination number $Z$, based on a jammed disordered packing~\cite{van2009jamming,liu2010jamming}, 
we designate two pairs of “source” and “target” nodes at distant sites. We selectively prune bonds until the ratio between strains on the source $\epsilon_S$ and target $\epsilon_T$, $\eta \equiv \epsilon_T/\epsilon_S$, first exceeds a goal value, $|\eta| \geq |\eta^*|$~\cite{rocks2017designing}. We analyze the evolution between function $A$, the target moving in-phase, to function $-A$, moving out-of-phase with the source.   

While different algorithms can be used to tune mechanical networks~\cite{rocks2017designing,yan2017architecture, hexner2021training, rouviere2023emergence}, 
here we use a greedy algorithm that minimizes a symmetric cost function based on induced stresses from network deformation due to strains applied at the source and target~\cite{pashine2021local}. Details of our tuning algorithm are described in Methods, Section~\ref{sec_methods}. 

We first tune a designated source and target pair for function $A$ where the target strain is in-phase with the source, such that $\eta_A \geq \eta^* > 0$. To achieve $-A$, we again start from the same initial network, and prune an independent set of bonds until $\eta_{-A} \leq -\eta^*$. This results in a source and target pair on two closely related networks, one of which can perform function $A$ and the other can perform function $-A$ depending on the set of bonds removed. By adding $(+)$ or subtracting $(-)$ a few bonds to a network tuned for one function, the network can switch to the other, as shown in Fig.~\ref{fig1_network}a. We refer to the bond modifications that separate the two networks as a set of mutations.
If a bond was pruned from the original network to create \textit{both} incompatible functions, it is not considered a mutation for this function switch.

We choose many pairs as source and target sites, and tune each of them as above for in-phase and out-of-phase functions. We separate this large ensemble of source-target pairs into groups with the same number of mutations, $M$. Figure~\ref{fig1_network}b shows $N(M)$, the distribution of $M$ obtained from one network with $\eta^* = 1$. We find similar distributions for pairs tuned using $\eta^* = 0.5, 2,$ and 4.

\begin{figure*}
\centering
\includegraphics[width=\linewidth]{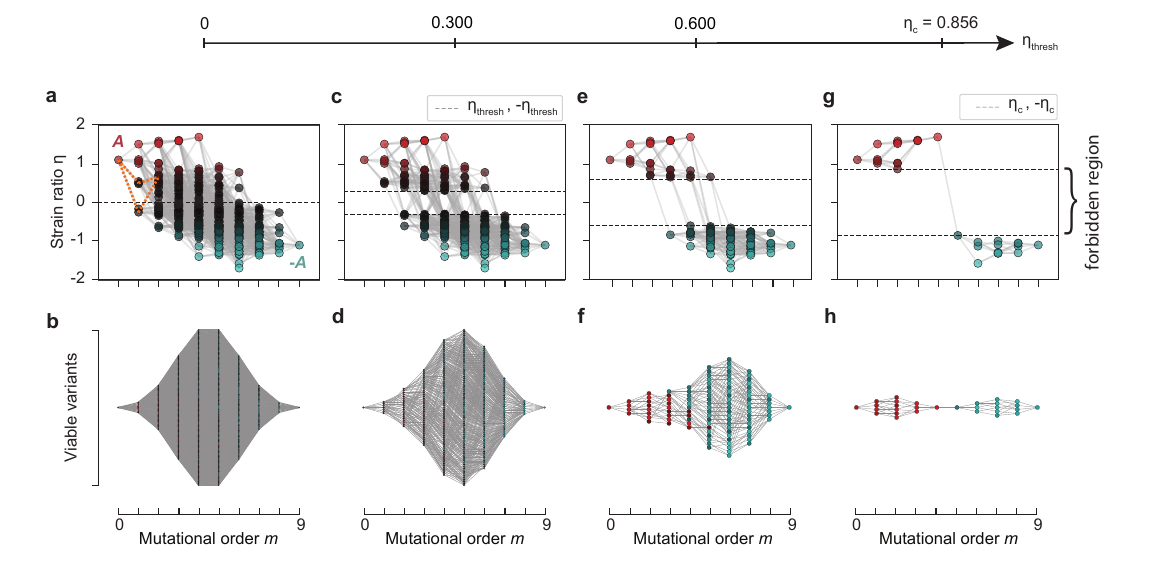}
\caption{
\textbf{Viable evolutionary paths with widening forbidden region.}
\textbf{Top row (a, c, e, g).} 
For a source-target pair from the $\eta^* = 1$ ensemble with $M=9$, viable pathways maintain single-step connections between the variants that fall outside a forbidden region $[-\eta_\text{thresh}, \eta_\text{thresh}]$ along a trajectory from $\eta_A$ at $m=0$ to $\eta_{-A}$ at $m=M$.  
Plots from left to right: $\eta_\text{thresh} = 0, 0.3, 0.6$, and $\eta_c = 0.856$, as indicated by the top arrow. Viable variants with $|\eta| > \eta_\text{thresh}$ are gradually eliminated as $\eta_\text{thresh}$ increases. Above a critical threshold $\eta_c$, no viable trajectories between $A$ and $-A$ exist.
\textbf{Bottom row (b, d, f, h).} Structures of viable evolution at each $\eta_\text{thresh}$. In all figures, variants with $\eta > \eta_\text{thresh}$ have function $A$ and are in blue, and variants with $\eta < -\eta_\text{thresh}$ have function $-A$ and are in red. The brightness of the colors scales with the magnitude of $\eta$.}
\label{fig2_varyingThresh}
\end{figure*}

\subsection{Creating an ensemble}

To evolve between the two networks with strain ratios  $\eta_A$ and $\eta_{-A}$, the system has to follow a trajectory of $M$ single-step mutations, which can occur in any order. Each mutation can be represented as a bit depending on whether it has occurred. The collective state of the system can be represented by an ordered binary sequence of $M$ mutations, resulting in $2^M$ possible combinations of mutations or “variants” and $M!$ possible evolutionary trajectories. This is a vast combinatorial space that we map out entirely for $M \leq 14$. Measuring the fitness $\eta$ for each combination of mutations (\textit{i.e.}, intermediate variant) provides complete fitness landscapes for large ensembles of function-switches, with statistics and evolutionary paths readily analyzed. This system is attractive in that there is a single definition of fitness that varies only by a sign between the two functions.


\begin{figure}[pht!]
\centering
\includegraphics[width=\linewidth]{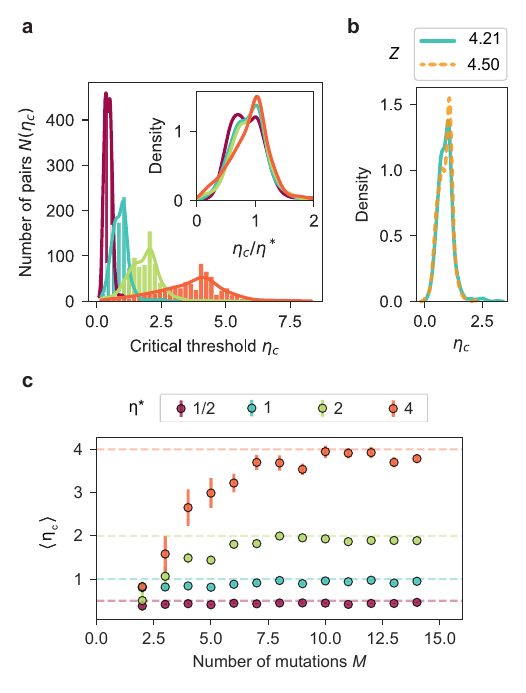}
\caption{\textbf{Critical fitness thresholds for viable evolution.} \textbf{a.} Distribution of critical thresholds for viable evolution $\eta_c$ for an ensemble generated from a network with $Z = 4.21$ using different initial tuning thresholds $\eta^*$ as shown in the legend. Inset: The same data plotted versus $\eta_c/\eta^*$. \textbf{b.} Comparison of ensembles generated from two networks, with $Z = 4.21$ and $Z = 4.50$ both tuned for $\eta^*=1$, show the $\eta_c$ distributions fall on top of each other. \textbf{c.} $\langle\eta_c\rangle$ versus  the number of mutations, $M$. The dashed lines show $\eta^*$ for each dataset with the same color.  }
\label{fig3_etac}
\end{figure}

\section{\label{results}Results}

\subsection{Network mutations are epistatic}

Figure~\ref{fig2_varyingThresh}a shows a complete fitness landscape with all possible trajectories for the function switch for one example source-target pair with $M=9$ tuned using $\eta^* = 1$ in a network with $Z=4.21$. The horizontal axis shows the order, or position in the trajectory, $m$, of the mutation: $m=0$ means no mutations, $m=1$ means one mutation (from any in our set of $M$), ... $m=M$ means all mutations have occurred.  The two endpoints were initially tuned for $A$ and $-A$: $\eta(m=0)\equiv \eta_A \geq \eta^* = 1$ and $\eta(m=M)\equiv \eta_{-A} \leq -\eta^* = -1$.

If there were no epistasis, the change in fitness $\eta$ due to a specific mutation would be independent of order $m$ at which it occurs. In Fig.~\ref{fig2_varyingThresh}a, this would result in parallel connections between neighboring columns (\textit{i.e.},  parallelograms between mutation pairs across three columns)~\cite{poelwijk2007empirical, bank2022epistasis}. However, as the dotted orange lines in the figure show, the structures between pairs of mutations can deviate strongly from parallelograms. Additionally, with no epistasis, the overall landscape between $A$ and $-A$ would be symmetric, but it is clearly not, which indicates strong epistasis in the network. 

Although epistatic interactions in biology have been attributed to many causes including competing molecular mechanisms (\textit{e.g.}, resistance, binding affinity, mechanical or thermodynamic stability)~\cite{depristo2005missense, poelwijk2007empirical} or biophysical pleiotropy~\cite{li2020biophysical}, we find that epistasis is more general. Epistatic interactions are strong in purely mechanical systems devoid of biochemical interactions.

\subsection{Varying viability threshold}

Living organisms evolve by incurring one mutation at a time. If a mutation results in the loss of biological function, the associated variant is deemed unfit and unlikely to survive natural selection~\cite{maynard1970natural}. Thus, if an organism loses one function during evolution, it must simultaneously acquire a different one to survive. This constraint implies a forbidden region, analogous to a fitness “valley of death” between $\eta_A$ and $\eta_{-A}$, that the organism must jump over without sacrificing fitness~\cite{wright1932roles, chiotti2014valley}. 

In our networks, the forbidden region corresponds to fitness being below a threshold value: $-\eta_\text{thresh} < \eta < \eta_\text{thresh}$. Any variant landing in the forbidden region is considered nonviable. The value of $\eta_\text{thresh}$ controls whether, and how easily, the network can evolve between $A$ and $-A$.  
Here we explore the extent to which evolutionary behavior is controlled by the threshold.

\subsubsection{The critical threshold}

Figure~\ref{fig2_varyingThresh} shows how the viable pathways in the fitness landscapes are modified by incrementally increasing $\eta_\text{thresh}$. For $\eta_\text{thresh} = 0$, all $2^M$ variants have $\abs{\eta} > 0$ so all trajectories are viable. 
As we increase $\eta_\text{thresh}$, pathways are progressively lost as more variants fall within the forbidden region as illustrated in the top row of Fig.~\ref{fig2_varyingThresh}; the graph becomes sparser as the gap expands. This persists until, at a critical value, the last viable pathway is reached at $\eta_\text{thresh} = \eta_c$. Above this value, there are no longer any viable paths.

Graphing the connected functional variants as a function of $m$ helps to visualize the global structure of the fitness landscape~\cite{Weinreich2006darwinian, steinberg2016environmental, poelwijk2019learning}.
The bottom row of Fig.~\ref{fig2_varyingThresh} shows this structure as $\eta_\text{thresh}$ is varied. The nodes in each column, plotted symmetrically around a horizontal axis, represent the viable variants at order $m$.
Each edge represents a single mutation connecting two variants at consecutive orders.

Figure~\ref{fig2_varyingThresh}b,d,f,h show the structure's progression as $\eta_\text{thresh}$ increases. The number of pathways decreases and the shape begins to form a bottleneck in the vicinity of the function switch.  At $\eta_\text{thresh} = \eta_c$ (Fig.~\ref{fig2_varyingThresh}h), only one “jumper” mutation can successfully cross the forbidden region. At that critical value,
the structure has the shape of a dumbbell whose thinnest part is a single line representing the lone surviving jumper mutation. 
Similar behavior is observed in proteins~\cite{anderson2015intermolecular,poelwijk2019learning}.

In the example shown in Fig.~\ref{fig2_varyingThresh}, the critical threshold of functional connectivity is $\eta_c = 0.856$.
It is nearly as large as the value for which the network was originally tuned: $\eta^* = 1$. This is not an aberration; we see this trend throughout our data.
Figure~\ref{fig3_etac}a shows the distributions of critical thresholds in four ensembles that were produced by tuning with $\eta^* = 0.5, 1, 2$ and 4. 
In the inset, those distributions plotted versus $\eta_c/\eta^*$ all peak near $\eta_c/\eta^* =1$.  To examine the effect of network geometry, we tuned a different network with higher coordination number (roughly twice as far from the jamming transition~\cite{van2009jamming,liu2010jamming}) $Z = 4.50$.   As shown in Fig.~\ref{fig3_etac}b, $\langle \eta_c\rangle/\eta^*\rightarrow 1$ independently of the coordination of the initial network.

Figure~\ref{fig3_etac}c shows the average $\langle\eta_c\rangle$ versus $M$ for those four ensembles.
In all cases, $\eta_c \rightarrow \eta^*$ for large enough $M$, indicating that, on average, there is a variant at each step along the trajectory that is nearly as fit as (or even more fit than) the initial tuned networks. 

The result $\eta_c \rightarrow \eta^*$ is surprising both because it is large and because its value is set by the original tuning protocol. One might have naively expected each mutation to contribute on average a change in fitness in a single step $\langle|\Delta \eta|\rangle \sim 2\eta^*/M$ with somewhat small fluctuations around that value. This is clearly not the case. On average, one single mutation, if at the right point in the trajectory, is able to create a jump in fitness $\Delta \eta \sim 2 \eta^*$.  This result is independent of ensemble coordination, $Z$, and (for large enough values) of $M$.  

\begin{figure}
\centering
\includegraphics[width=\linewidth]{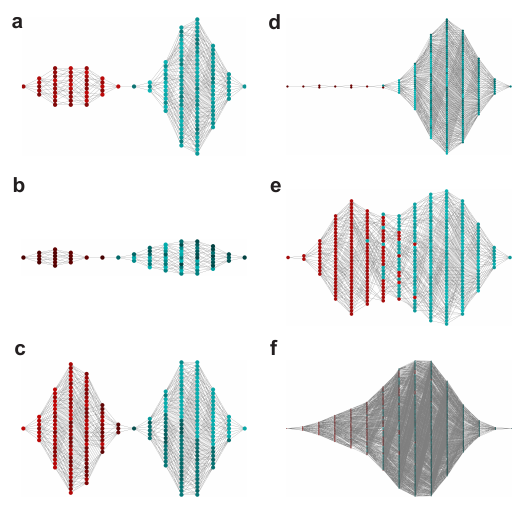}
\caption{\textbf{Functional connectivity structures at critical threshold.} Examples of six structures at the threshold of functional connectivity from source-target pairs tuned using $\eta^* = 1$ with $M=14$. Red variants are for $A$, blue for $-A$, and the brightness of each color scales with fitness magnitude $|\eta|$, normalized by the largest fitness of all variants for each pair. 
\textbf{a., b.} Examples of structures with a dumbbell configuration where function switch is connected by a single thread.
\textbf{c.} Two clusters with only a single variant at the edge of the forbidden region, connected to many others across the barrier. 
\textbf{d., e.} Function-switch connections at multiple orders. 
\textbf{f.} Trajectory endpoints connected by multiple pathways with no narrow neck at any intermediate order.}

\label{fig4_topologies}
\end{figure}

\subsubsection{Structure of viable pathways}\label{sec_results_struct}

The dumbbell shape in Fig.~\ref{fig2_varyingThresh}h shows that many pathways exist between variants with the same function, yet
only one mutation connects those with opposite function. The function switch is most likely to occur near $M/2$ where the number of variants is largest.  
To assess the commonality of this behavior, we examine the variety of structures found at $\eta_c$ in our ensembles. Figure~\ref{fig4_topologies} shows archetypes of the six most common shapes. 
 
In all the examples in the first column of Fig.~\ref{fig4_topologies}, the neck shrinks either to a single line or one node at the function switch.   
This occurs when one variant essential for the switch lies at $\eta_c$. We find this behavior and corresponding structures in $\sim 60\%$ of the source-target pairs in the $\eta^* = 1$ ensemble.

In the remaining shapes, shown in the second column of Fig.~\ref{fig4_topologies}, multiple pathways cross the forbidden region. In these cases, the threshold is determined by a variant that is not contiguous to where the function switch occurs. 
To illuminate this, we compare Fig.~\ref{fig4_topologies}c and d:   
Fig.~\ref{fig4_topologies}c shows only a single variant with $-A$ sits at the boundary of the forbidden region; this creates an extreme bottleneck for evolution. In contrast, Fig.~\ref{fig4_topologies}d shows two different jumps at $m=5$ and $m=6$ while $\eta_c$ is determined by the variant at $m=0$. The bottleneck broadens further in Fig.~\ref{fig4_topologies}e, and disappears completely in Fig.~\ref{fig4_topologies}f, where the shape is reminiscent of Fig.~\ref{fig2_varyingThresh}b where $\eta_\text{thresh} = 0$.

\subsubsection{Bounds for fitness at the function switch}\label{sec:alt_etac}

\begin{figure*}
\centering
\includegraphics[width=\linewidth]{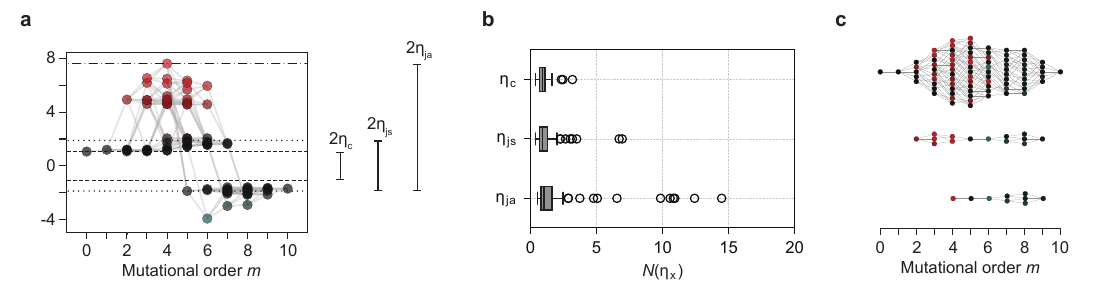}
\caption{\textbf{Upper bounds on fitness.} \textbf{a.} Different measures of fitness thresholds determined strictly by the largest function-switch “jump”. $\eta_{js}$ defines a threshold that is still symmetric about $\eta = 0$; $\eta_{ja}$ allows for asymmetry, representing the highest possible fitness for function switch. \textbf{b.} Statistics describing the three measures for $\eta^*=1$ data with $M=10$. \textbf{c.} Structures of viable pathways corresponding to each measure.}
\label{fig5_diffdefs}
\end{figure*}

As shown above, it is possible for the network to increase its fitness \textit{en route} to achieving the opposite function. How high can the fitness for this switch be? As shown in Fig.~\ref{fig5_diffdefs}a, $\eta_c$ is set by the tuned network fitness $\eta_A$ at $m=0$, even though the function-switch mutations are considerably more fit.  
Here we consider only those variants connected to the jumper that are at least as fit as the jumper itself. This leads to a larger bound on fitness than $\eta_c$ and shorter trajectories between variants with functions $A$ and $-A$. We define a larger fitness threshold $\eta_{js}$ that is determined by fittest jumper, but still symmetric about $\eta=0$.
A third measure is to include the full extent of the change in fitness $\Delta\eta$ between any two variants with opposite functions, $\eta_{ja}$, regardless of symmetry around zero. This is guaranteed to produce thresholds with $\eta_{ja} \geq \eta_{js} \geq \eta_c$. These definitions are illustrated in Fig.~\ref{fig5_diffdefs}a.

The statistics for these three measures of viable pathways between the opposite functions -- $\eta_c$, $\eta_{js}$ and $\eta_{ja}$ -- are shown in Fig.~\ref{fig5_diffdefs}b. The distribution of viable pathways at the corresponding fitness thresholds are qualitatively similar but $\eta_{js}$ and $\eta_{ja}$ have higher  averages with longer tails to higher $\eta$. This highlights the dramatically large changes in function fitness that are possible due a single mutation. In some cases shown, the mutations create jumps in the fitness that are more than $14\eta^*$.

Figure~\ref{fig5_diffdefs}c shows the corresponding functional connectivity structures for all three cases. When $\eta_c$ defines the  threshold, the viable fitness landscape structure appears with no thin-neck region similar to the archetypal shape in Fig.~\ref{fig4_topologies}f. By expanding the threshold beyond $\eta_c$, the same fitness landscape can be represented with a dumbbell structure similar to those in the left-hand column of Fig.\ref{fig4_topologies}. 

\subsection{Distinct classes of mutations}\label{sec_results_locmut}

As we emphasized, the impact of a given mutation depends on the specific set of mutations that have already occurred. Here we show that the average effect of a mutation, $\langle \Delta \eta_i \rangle$, depends on its order $m$ along the trajectory.  

\begin{figure}
\centering
\includegraphics[width=\linewidth]{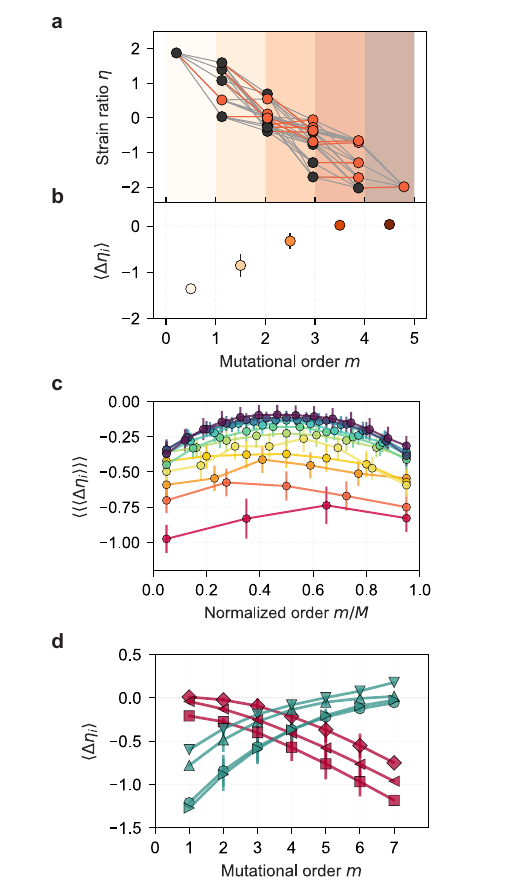}
\caption{\textbf{Average mutational effects versus order.}
\textbf{a.} Fitness landscape for a source-target pair with $M=5$; the orange lines show all the  changes in fitness due to one of the five mutations.
\textbf{b.} Highlighted data in (\textbf{a}) averaged over each column, $\langle \Delta\eta_i\rangle$, versus $m$. 
\textbf{c.} The change in fitness averaged over entries in each column, over all mutations $i$ and over all members of the $\eta^* = 1$ ensemble, $\langle\langle\langle\Delta\eta_i\rangle\rangle\rangle$, versus $m/M$. Each curve represents data for a given $M$ with $4 \leq M \leq 14$. 
\textbf{d.} Averaged curves, as in (\textbf{b}), for all mutations for an example source-target pair with $M=7$ illustrating two distinct epistatic classes with opposite behaviors.}
\label{fig6_locmut}
\end{figure}

Figure~\ref{fig6_locmut}a shows a graphically modified fitness landscape for a network with $M=5$ where we have highlighted in orange all the connections, along with the variants themselves, where a specific mutation, $i$, occurs. The slopes of these lines, $\Delta \eta_i/\Delta m= \Delta \eta_i$ are the local contributions of mutation $i$. By looking at the average slopes of the orange lines as a function of order $m$, we see that the average effect of mutation $i$, $\langle \Delta \eta_i \rangle$ varies with $m$ as shown explicitly in Fig.~\ref{fig6_locmut}b. We can do this same averaging for the other $(M-1)$ mutations. We note that if there were no epistasis, all these lines would have been parallel across all $m$.

Figure~\ref{fig6_locmut}c shows $\langle\langle\langle \Delta\eta_i \rangle\rangle\rangle$ versus $m/M$ where $\langle\langle\langle ... \rangle\rangle\rangle$ indicates an average over all contributions of a mutation at a given order (\textit{i.e.}, in a column), over all mutations for that source/target pair, and over all source/target pairs in the ensemble with a given value of $M$.  The average effects of mutations are largest at $m=0$ and $m=M$ and smallest at the midpoint near $M/2$. While we might be led to expect the largest mutational effect to be at $\sim M/2$ where the function switch at $\eta_c$ is most likely to occur, the curve in Fig.~\ref{fig6_locmut}c counter-intuitively shows the changes are smallest near that value.

This apparent discrepancy is clarified in Fig.~\ref{fig6_locmut}d, which illustrates the average behavior of all individual mutations $\langle \Delta \eta_i \rangle$ versus $m$ for one example source-target pair. This example reveals two distinct classes of mutations: one, plotted in blue, where the effect is large, $\Delta \eta \sim \eta^*$, near $m=0$ and approaches zero near $m=M$; the other, plotted in red, with the inverse behavior. Most of the mutations in our ensemble have this feature; for source-target pairs with $M \geq 8$ in the $\eta^* = 1$ ensemble, more than $80\%$ of mutations are in one of these two classes. 

The parabolic shape in Fig.~\ref{fig6_locmut}b emerges by averaging the two classes of mutations with opposite trends. 
Near $m=M/2$, where the function switch is most likely to occur, the effects from both classes are comparable.  

To have a large critical threshold, the single-step jump in fitness, $\Delta \eta$, must be large and optimally located so that it traverses the forbidden region that is symmetric around $\eta = 0$. 
Moreover, the remaining mutations should not be deleterious so that the remainder of the trajectory lies above the threshold. 
Ideally, one extremely large jump would be accompanied by $(M-1)$ mutations with negligible $\Delta \eta$ as illustrated in Fig.~\ref{fig2_varyingThresh}. Starting from either $m=0$ or $m=M$, the first mutations should cause only small changes in fitness, leaving $\eta$ close to or above that of the starting point.   If one had instead chosen as an initial mutation the one with a large $\Delta \eta$, this would have had $\Delta \eta \sim \eta^*$ and left the network with a fitness near $\eta = 0$ inside the forbidden region. (To achieve $\eta_c \sim \eta^*$ requires instead $\Delta \eta \sim 2\eta_c$.) Near $m\sim M/2$, the variance is large and can produce a large $\Delta \eta$.  

\section{Discussion}

Similarities exist between our results and those reported in proteins.  Novel features also emerge.

\subsubsection{Memory of a common ancestor, vestigial mutations, and enhanced adaptability}

We have shown in Fig.~\ref{fig6_locmut} two distinct classes of mutations each independently relevant to one of the two incompatible functions. We find that the set of mutations in the class relevant to $A$ (and similarly to $-A$) corresponds exactly to the set of bonds pruned to obtain that function from the initial common network. Thus, if we had begun our analysis with no prior knowledge of the initial network, we could still retrieve it (with the exception of any bonds that had to be pruned to obtain \textit{both} functions) by inspecting the two distinct epistatic classes. In other words, the two classes leave an imprint or memory of the common ancestor from which the two fit variants evolved. 

Not all $M$ mutations are necessary for a function-switch. As shown in Fig.~\ref{fig5_diffdefs}, one need not be at a trajectory endpoint to have sufficient function fitness -- there are fit (and possibly fitter) variants elsewhere along the trajectory. Thus, we can obtain one function while leaving behind a few mutations beneficial for the opposite one. Starting with one of these intermediate variants would result in vestigial remnants of the original evolutionary pathway and lead to a more adaptable network. 

It was previously shown that adaptability can be enhanced by varying the background environment during training~\cite{falk2023learning}. Having identified an optimal trajectory between the desired functions -- specifically a sole mutation between two variants that are even more fit than those for which the system was originally trained (such as those identified in Fig.~\ref{fig5_diffdefs}) -- we have also created an exquisitely adaptable network. These results show that we can reliably find the \textit{single} mutation that produces the function switch. This may be relevant for creating adaptable networks in materials science.
Additionally, for a designated fitness threshold, Fig.~\ref{fig4_topologies} shows that many paths allow switching between opposite functions at different mutational orders.


\subsubsection{Viable pathways in networks and proteins}


Dumbbell structures, such as those in the left column of Fig.~\ref{fig4_topologies}, are also observed in protein evolution where it was asserted that these shapes, important for understanding the relationship of epistasis to evolvability, demonstrated severe epistatic constraints~\cite{poelwijk2019learning}. However, the thin necks that we analyze are not necessarily related to epistatic effects, nor do they always coincide with the function switch. They emerge simply due to nature of the critical threshold when the final remaining mutation that can switch function is the one that sets the maximum threshold value. While the characteristic structures illustrated in Fig.~\ref{fig4_topologies} were analyzed by focusing on trajectories remaining at the \textit{critical} thresholds, our results should still be applicable to the structures found in proteins that may be considered as perturbations around $\eta_c$.

We demonstrate in Fig.~\ref{fig5_diffdefs} how simply changing where the trajectory begins or ends -- without altering the epistatic interactions -- can change the shape from one that resembles Fig.~\ref{fig4_topologies}f to the dumbbell shape of Fig.~\ref{fig4_topologies}a.
While extreme values may be enhanced due to epistatic effects, we find many cases with large critical thresholds even without epistasis as discussed below and in the SI.

The data in Fig.~\ref{fig6_locmut}c of $\langle\langle\langle \Delta \eta \rangle\rangle\rangle$ versus order, $m/M$ are consistent with trends in protein evolution, where the fitness changes $\Delta \eta$, decrease with more mutations or in fitter backgrounds.  These are typically attributed to diminishing effects of epistasis and have been modeled for biological evolution~\cite{bershtein2006robustness, barrick2009genome, maclean2010diminishing, chou2011diminishing, couce2015rule,reddy2021global}.
Our results, in contrast, suggest that this behavior is not specific to living organisms but emerges in mechanical networks by averaging the effects of two epistatic classes.

\subsubsection{Evolutionary structure}
 
The effect of an individual mutation, such as the one highlighted in Fig.~\ref{fig6_locmut}a, can have very different effects depending on the other mutations present; a given mutation can in some cases increase fitness and in other cases decrease it even at the same value of $m$.  This is reminiscent of idiosyncratic epistasis observed in biological evolution~\cite{lyons2020idiosyncratic, reddy2021global}.  

Despite this variability and context-dependence, there are systematic variations in the behavior, as demonstrated in Fig.~\ref{fig6_locmut}d when suitable averages are taken.  The effects of each mutation is not completely random; in some cases simply knowing in which of the two classes a mutation belongs provides predictive capability of its performance at different orders. This effect of epistasis causes a common variation of $\Delta\eta_i$ with $m$ over all the members in each class. By attributing specific mutations to their relevant functional classes, we can clarify their systematic roles in the evolution, potentially making evolution more predictable and adaptation more efficient.


\subsubsection{Necessity of epistatic effects}

It is not absolutely necessary for an evolutionary trajectory to be epistatic in order for viable pathways to exist between two functions~\cite{mccandlish2013role}. 
Provided the contributions from the different mutations have a broad enough spread and occur in the optimal order, systems with context-independent (\textit{i.e.}, non-epistatic) effects can also lead to \textit{large} thresholds. In fact, in the example shown in Fig.~\ref{fig6_locmut}d, if the first-order effects obtained at $m=0$ were constant throughout the trajectory (that is, the values for each $\Delta \eta_i$ were taken from the first column), the critical threshold would even be marginally larger than the measured threshold with epistasis, $0.603 > \eta_c = 0.597$. 

While individual examples show that systems without epistasis can have large critical thresholds, on average the critical threshold values do decrease as $M$ increases. Further examples and statistics are provided in the SI.

\section{Conclusions}

The ease of tuning our networks to incorporate function allows the creation of large ensembles with which to probe the evolution of matter. Even with the advent of technologies such as deep mutational scans~\cite{fowler2014deep}, these can be difficult to obtain in biological experiments. These physical systems allow us to study structure-to-function maps, evolution and epistasis devoid from the complexities of biochemical interactions. As we have demonstrated, there is a generality that extends beyond biological organisms; we find many similarities between the evolution seen in proteins and what is found in our system. This suggests that there may be a common underlying explanation of these behaviors that is captured in the simplicity of mechanical networks.

The importance of the order in which interactions occur is not commonly considered in condensed matter, despite there being systems where epistasis naturally occurs, such as in the study of meta-materials, where origami (folding of sheets) is sensitive to the order of the folds~\cite{stern2017complexity, jules2022delicate}, and in memory formation, where interacting hysterons depend on the training sequence~\cite{van2021profusion, lindeman2021multiple}. Our networks, regarded as a platform on which to study evolution, allow some of the questions addressed in this paper to be asked of those systems. It paves the way for new questions to be formulated about the importance of sequence of interactions and time-ordering of events.

There are many directions for future work. In particular, the two starting functions $A$ and $-A$ could have been obtained in different ways from the original network. It is important to understand the  extent to which our findings depend on our tuning algorithm.  For example, one could ask whether different algorithms produce different kinds of memories of their ancestry. This work also opens the door to studying the evolution of networks in other physically relevant contexts by including the effects of pre-stress and nonlinear elastic response. The definition of fitness could also be augmented to include more than one function,  which allows us to test our findings against different models of molecular- and macro- evolution and susceptibility to an environmental selection pressure.
We hope to probe the cause for the large function switch and uncover aspects of the mechanical basis of evolution by studying soft normal modes. Finally, while we have concentrated on studying the average values for the entire ensemble, the question remains as to whether all members of the ensemble -- for example, systems with large and small critical thresholds -- behave similarly.


While we see a generality between proteins and networks, there may be more than a single form of evolutionary behavior with other characteristics. It would be exciting to explore the extent of this generality and search for other forms of evolvable matter.


\section{Methods}\label{sec_methods}

\subsection*{Tuning-by-pruning algorithm}

We choose a pair of source nodes and a pair of target nodes such that, initially, neither pair is  directly connected by a bond. 
We apply a strain between the two source nodes or between the two target nodes along the line connecting them as shown in Fig.~\ref{fig1_network}a.  Any deformation on the network, applied either at the source or at the target, pushes it out of equilibrium and generates forces -- tension or compression -- on the other bonds in the network. To tune for in-phase behavior $A$, we apply stresses at the source and measure the stress on each bond in the network. Independently, we apply a strain only to the target nodes that is in-phase with how the strain had been applied to the source. We calculate the product of the stresses on each bond due to the two separate perturbations.   

In the first iteration, the algorithm removes the bond in the network that has the maximum product of stresses from the source and target. We measure $\eta$ after the bond is removed and repeat this operation multiple times until we reach $\eta \geq \eta^*$  where $\eta^*$ is the minimum desired response. We use the same algorithm to tune for out-of-phase behavior $-A$ by applying source and target stresses that are out of phase with each other.  We continue until $\eta \leq -\eta^*$~\cite{pashine2021local}.  Because we only tune for the two responses to be greater than a target value, $|\eta^*|$, the value of $\eta$ found for the in-phase function may not be exactly equal to the one found for the out-of-phase behavior.

\begin{acknowledgments}

We are grateful to Martin Falk, Chloe Lindeman, Thierry Mora, Arvind Murugan, Rama Ranganathan, Eric Rouviere, Aleksandra Walczak, and Francesco Zamponi for enlightening discussions.  We thank Ayanna Matthews and Nidhi Pashine for providing networks derived from jammed packings and for procedures to tune them.   This work was supported by DOE Basic Energy Sciences Grant DE-SC0020972. SA received support from NSF Center for Living Systems grant 2317138.

\end{acknowledgments}

\bibliography{bibli}

\newpage
\newpage
\newpage

\appendix

\clearpage
\onecolumngrid

\section*{Supplementary Information}

\subsection*{Non-epistatic models}\label{secA1}

As shown in Fig.~\ref{fig6_locmut} in the main text, epistasis creates a dependence of a mutation's effect on the order at which it occurs $m$; in the absence of epistasis, the contribution of each mutation would be constant across all $m$. The key issue in comparing our epistatic results to equivalent cases without epistasis is the critical choice of $\Delta \eta_i$ -- determining a single value that would represent the \textit{constant} additive change to $\eta$ induced by an individual mutation $i$. Since no single constant $\Delta\eta_i$ can be directly measured from the data, this parameter can only be determined from an additive, “non-epistatic” model, making the comparison with our espitatic results inherently model-dependent. 

To assess the necessity of epistasis to obtaining the large threshold of viable evolution, we reconstruct the fitness landscapes in the $\eta^*=1$ ensemble according to three different models of strictly additive effects of mutations, using different measures for a constant $\Delta_i$: (1) first-order contribution relative to $\eta_A$, (2) re-scaled first-order effect, and (3) re-scaled average effect.

\begin{figure}[b]
    \centering
    \includegraphics[width=0.65\linewidth]{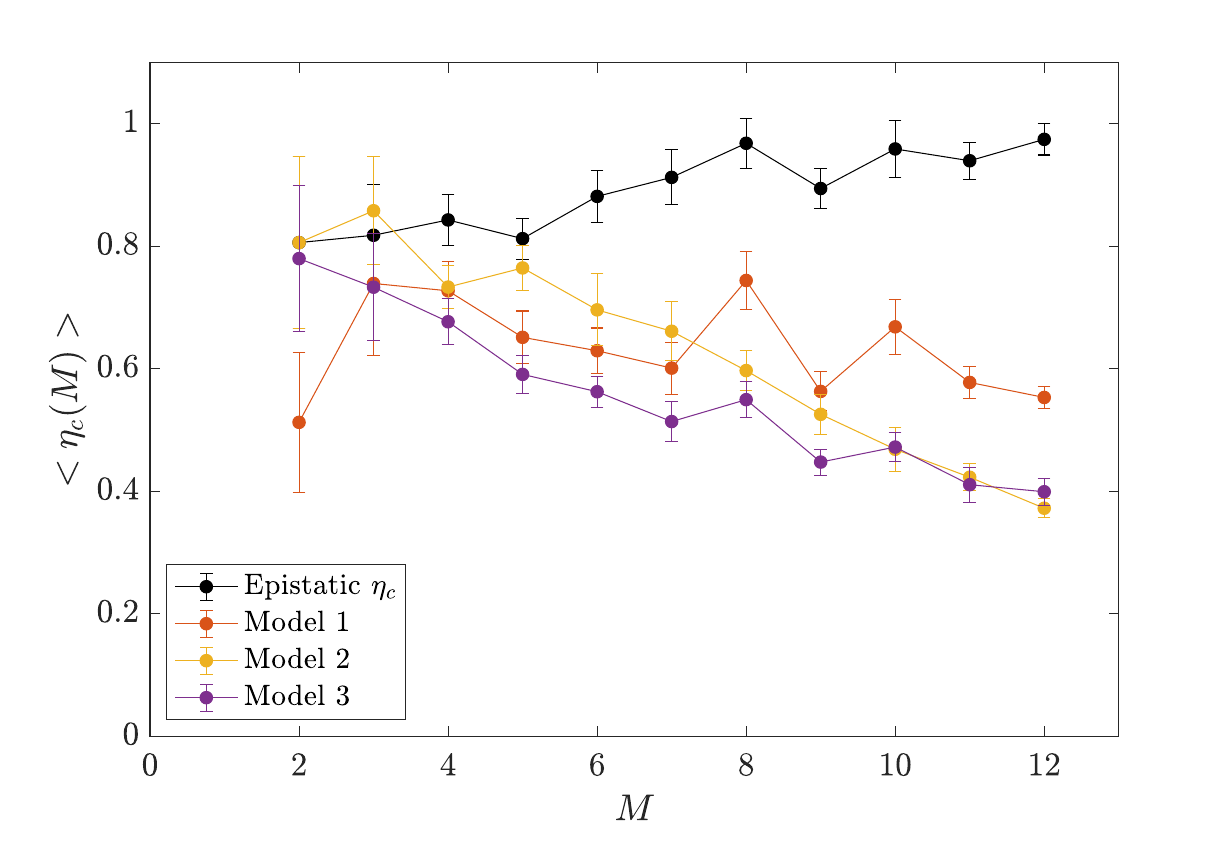}
    \caption{Average critical thresholds $\langle\eta_c(M)\rangle$ using non-epistatic reconstructions of the data in the $\eta^*=1$ ensemble. The different models are shown in the legend with the black data points corresponding to the data for the original epistatic ensemble.}
    \label{fig:enter-label}
\end{figure}

\paragraph*{\newline}
\textit{Model 1.} The first model assumes non-diminishing effects, such that the constant effect of a mutation relative to any intermediate variant does not change from that relative to $\eta_A$. The contribution due to a mutation $i$, across all $m$, is its first order contribution: 
\begin{equation}\label{eq_firstOrder}
    \Delta \eta_{i} = \eta_{i}^{(m=1)} - \eta_A
\end{equation}

A source-target pair connected by $M$ mutations will have a set of $M$ constant contributions due to each distinct mutation. 
To reconstruct a non-epistatic fitness landscape for the pair, we start from its measured parental variant $\eta_A$ from the as-tuned network. We then measure the first-order contribution due to each mutation as in Eq.~\ref{eq_firstOrder} above. Then the fitness of each non-epistatic variant $\overline{\eta}_{\alpha}$ (where $\alpha$ specifies one of the $2^M$ variants comprising the fitness landscape) is the sum of the contribution of its comprising mutations:
\begin{equation}
    \overline{\eta}_{\alpha} = \eta_A + \sum_i \Delta\eta_{i}
\end{equation}
where the sum is over all mutations, $i$, that specify $\alpha$.

\paragraph*{\newline}
\textit{Model 2.} As shown in Fig. 6 in the main text, first-order contributions likely over-estimate the cumulative effects of mutations and adding them can overshoot trajectory endpoint $\eta_{-A}$ (that is, $\overline{\eta}_{-A} > \eta_{-A}$). Therefore, in the second model, we re-scale the first-order effects to impose the constraint that they sum up to the difference between the parental variants. This ensures that the ends of the trajectories remain the same in the epistatic and non-epistatic scenarios. 
To do this, we rescale all the individual mutation by a constant factor $c$ such that, 
\begin{equation}
   \eta_A + c\sum_{i}^{M} \Delta \eta_{i}^{(m=1)} = \eta_{-A}
\end{equation}
then re-scale the contribution of first-order effects in reconstructing the non-epistatic fitness landscape, 
\begin{equation}
    \overline{\eta}_{\alpha} = \eta_A + c\sum_j \Delta\eta_{j}
\end{equation}
where the sum is again over mutations that specify $\alpha$.

\paragraph*{\newline}
\textit{Model 3.} Here we take the average effect of a mutation $i$ across all $m$, irrespective of the preceding variant. This would be equivalent to taking an average of all $2^{M-1}$ slopes in Fig.6a, $\langle\langle \Delta\eta_i \rangle\rangle$. Similarly to model 2, we re-scale this average contribution such that the total contribution produces the same trajectory endpoints as in the epistatic landscape. The reconstructed variants are given by
\begin{equation}
    \overline{\eta}_\alpha = c\sum_i \langle\langle \Delta\eta_i \rangle\rangle
\end{equation}
where the re-scale factor $c = (\eta_A - \eta_{-A})/\sum_i \langle\langle \Delta\eta_i \rangle\rangle$.

\paragraph*{\newline}
The average critical threshold, $\langle 
\eta_c \rangle$,  averaged over the entire $\eta^* = 1$ ensemble, is shown for the three models in the Fig. 1 in the SI as a function of $M$.  For comparison, the black data points indicate the value of the average critical threshold in the epistatic landscape as shown in Fig.~3 in the main text. In the models with no epistasis, the critical threshold, on average, decreases with $M$ and can drop to $\approx 0.4$ at $M=12$. While the trends for the critical threshold do drop, their average decrease is only by a factor of ~2.  Moreover, we have found that in many individual cases, the value for $\langle 
\eta_c \rangle$ in these models leads to a larger critical threshold than that for the original epistatic case.

\end{document}